\documentclass[amsmath, amsfonts, showpacs, superscriptaddress, twocolumn, prl]{revtex4}
\usepackage{graphicx,braket}
\usepackage{epsfig}
\usepackage{bm}
\usepackage{dcolumn}
\usepackage{amsmath}
\usepackage{amssymb}

\begin{document}

\title{Brownian scattering of a spinon in a Luttinger liquid}

\author{M.-T.~Rieder}
\affiliation{Dahlem Center for Complex Quantum Systems and Institut f\"ur Theoretische Physik, Freie Universit\"at Berlin, 14195 Berlin, Germany}

\author{A.~Levchenko}
\affiliation{Department of Physics and Astronomy, Michigan State University, East Lansing, Michigan 48824, USA}

\author{T.~Micklitz}
\affiliation{Centro Brasileiro de Pesquisas F\'isicas, Rua Xavier Sigaud 150, 22290-180, Rio de Janeiro, Brazil}

\begin{abstract}
We consider strongly interacting one-dimensional electron liquids where elementary excitations carry either spin or charge. At small temperatures a spinon created at the bottom of its band scatters off low-energy spin- and charge-excitations and follows the diffusive motion of a Brownian particle in momentum space. We calculate the mobility characterizing these processes, and show that the resulting diffusion coefficient of the spinon is parametrically enhanced at low temperatures compared to that of a mobile impurity in a spinless Luttinger liquid. We briefly discuss that this hints at the relevance of spin in the process of equilibration of strongly interacting one-dimensional electrons, and comment on implications for transport in clean single channel quantum wires.
\end{abstract}

\date{December 20, 2014}

\pacs{71.10.Pm, 72.10.-d, 73.63.Nm, 05.40.Jc}

\maketitle

{\it Introduction.} The Luttinger liquid model of interacting one-dimensional (1D) electrons advocates that  spin and charge degrees of freedom of electrons deconfine into elementary excitations which represent collective waves of spin and charge density~\cite{Stone,Gogolin,Giamarchi}. These collective bosonic modes do not interact and propagate with different velocities so that the charge and spin of an electron move apart in time~\cite{DL,Luther-Emery,Ogata-Shiba}. This scenario of spin-charge separation is a remarkable example of the fractionalization of a quantum number occurring when low energy quasiparticles of a strongly interacting system do not evince much resemblance to the underlying electrons. Indeed, the subsequent refermionization of the bosonized model reveals that elementary excitations are charged spinless quasiparticles  -- holons, and neutral spin-1/2 quasiparticles -- spinons. Of course, this paradigm of spin-charge separation is an idealization which is violated once effects of band curvature are accounted for~\cite{Brazovskii-JETPLett93,Nayak-PRB01}. In general, spin-charge coupling leads to a plethora of pronounced effects and the properties of nonlinear Luttinger liquids are currently under intensive investigation~\cite{Review-1,Review-2}. 

At small but finite spin-charge coupling spinons are the lowest energy excitations of the strongly interacting one-dimensional system at any given momentum. While holons readily decay into a continuum of spin-excitations even at zero temperature~\cite{Schmidt-PRB10,Pereira-Sela-PRB10}, the spinon remains a stable quasiparticle. That is, creating a hole with large momentum (e.g. close to the spinon's  band bottom) in a strongly interacting one-dimensional electron liquid, the charge is rapidly screened leaving behind the neutral spin-$1/2$ fermionic degree of freedom. At finite but small temperatures one may then study the dynamics of such a large-momentum spinon.

Knowledge of the spinon's dynamics is of fundamental interest, as it is the backscattering of large-momentum  spinons which eventually equilibrates one-dimensional electronic liquids. Indeed, the relaxation of strongly interacting electrons occurs as a result of a multi-stage process~\cite{AL-K,Matveev-Review}. In the first stage, excitations scatter of each other and relax to a common equilibrium. The generic equilibrium of the liquid of excitations is characterized by a temperature, reflecting energy-conservation, and a drift velocity, accounting for a finite motion of a system with momentum-conservation. A finite coupling between spin and charge modes allows for an exchange of energy and momentum and both subsystems are then characterized by a common temperature and drift velocity. The rate of such prethermalization is relatively fast and follows power-law temperature dependence $\propto T^3$~\cite{AL}.
 
In the second stage, the total momentum of excitations relaxes. The relevant processes can be viewed as the umklapp scattering of large-momentum bosonic excitations in which momentum is transfered to the zero modes of the Luttinger liquid~\cite{Matveev-Andreev-PRL11}. Alternatively, it can be described by backscattering of a fermionic excitations~\cite{RMM-PRL09}, and we use both languages interchangeably throughout this paper. At strong interactions the key process for equilibration is the backscattering of large-momentum spinons. That is, spin excitations first relax their drift velocity and then drag the fluid of charge excitations to equilibrium. As a result, the rate of full equilibration is slow and follows the activated temperature dependence $\propto e^{-\Delta_\sigma/T}$ with $\Delta_\sigma$ being the width of the spinon band. 

Given the prominent role of spinons close to the band bottom in the process of equilibration it is interesting to study their kinetics. Previous work~\cite{Neto-Fisher} has studied the dynamics of a mobile impurity in a spinless Luttinger liquid (which is  also deeply related to the problem of dark soliton decay in a Bose systems~\cite{Gangardt-Kamenev,Schecter}). Regarding the spinon at the bottom of the band as a mobile impurity, we will closely follow the approach of Ref.~\cite{Neto-Fisher} and include spin into the picture. In a similar spirit, the recent study Ref.~\cite{Lamacraft} considered a spin-1/2 impurity coupled antiferromagnetically to a one-dimensional gas of electrons and showed the formation of an unconventional Kondo effect. Here we study the complementary case of a ferromagnetic coupling where the impurity-spin remains unscreened. We find that, similar to the antiferromagnetic case, the mobility of the spinon is parametrically suppressed at low temperatures as compared to the impurity diffusion in a spinless Luttinger liquid. However, unlike the antiferromegnatic case, mobility ceases to be universal in the low temperature limit and strongly depends on the integrability breaking perturbations of the Luttinger liquid model. Suppression of the mobility hints at the relevance of the spin degree of freedom for equilibration of strongly interacting one-dimensional electrons, and we briefly comment on its implications for the conductance in quantum wires. Similar findings have been reported in a recent elegant work where a particular realization of the Luttinger liquid model -- a Wigner crystal at low electron density -- was considered~\cite{MAK-PRB14}. 

{\it Fokker-Planck equation and diffusion constant}. The lowest energy excitation for a system with concave spectrum at a given momentum $p$ is a hole. For weak interactions the hole carries spin and charge quantum numbers of the removed electron. In a strongly interacting electron fluid, on the other hand, the charge is screened and a neutral spin-1/2 particle, the spinon, remains with a two-fold degeneracy of the energy levels protected by spin-rotation symmetry. In the following we assume that a spinon in the vicinity of the band bottom has a quadratic dispersion
\begin{equation}\label{disp}
\varepsilon^\sigma_p=\Delta_\sigma - {p^2\over 2m^*},
\end{equation} 
with $m^*$ being the effective mass. At finite temperature $T$ the typical momentum of a spinon is of the order 
$p\sim\sqrt{m^*T}$ and should be compared to the momentum exchanged in collisions with low energy charge- or spin-excitations $\delta p\sim T/v_{\rho(\sigma)}$, where $v_{\rho(\sigma)}$ are their respective velocities. 
If temperatures are sufficiently low, $T\ll m^*v_{\rho(\sigma)}$, relative changes are small $\delta p/p\ll1$ and the spinon may be viewed as a heavy Brownian particle propagating in a gas of light particles, viz. the low energy excitations. Collisions with the light particles renders the motion of the spinon to be diffusive in momentum space, and the kinetic equation describing the spinon distribution $F(p)$ is approximated by the Fokker-Planck form (throughout the paper we set $\hbar=k_B=1$) 
\begin{equation}\label{fp}
\partial_t F(p)={D\over 2} \partial_p \left(-{p\over m^* T}+\partial_p\right)F(p).
\end{equation}
Here we employed that by spin rotation symmetry $F(p)$ is independent of the spin orientation, 
and for simplicity we will concentrate on homogeneous liquids. The microscopics of this Brownian motion in momentum space is governed by the diffusion constant
\begin{eqnarray}\label{d}
&&\hskip-.45cm
D(T)=\sum_q q^2 W(q),
\nonumber \\
&&\hskip-.45cm
W(q)=\sum_{ ll'}\sum_{\varsigma\varsigma',kk'}
\Gamma^{\varsigma\varsigma'}_{ll'}(q;kk')n_l(k)[1+n_{l'}(k')].\label{W}
\end{eqnarray}
 Here $W(q)$ is the probability for a collision in which the spinon changes its momentum by $q$ and we already anticipated that at low momenta of interest it only depends on the transferred momentum. Scattering rates $\Gamma^{\varsigma\varsigma'}_{ll'}(q;kk')$ describe processes in which a spinon with momentum $p$ and spin-projection value $\varsigma$ is scattered into a state with ${p+q,\varsigma'}$ by absorbing a spin (charge) excitation $l= \sigma(\rho)$ with momentum $k$ and emitting an $l'$-excitation with momentum $k'=k-q$. The bosonic occupation numbers for charge (spin) excitations at momentum $k$ are $n_{\rho(\sigma)}(k)$ and we have set the occupation of a missing large-momentum spinon $1-F(p+q)\simeq 1$. Finally, the mobility of the spinon is related to the diffusion constant by the usual kinetic formula $\mu(T)=T/D(T)$.
 
Consequences of the Fokker-Planck equation on 1D electronic transport are well studied in the literature \cite{TJK-PRB10,Matveev2012,DGP,Rieder2014}. Here, our main interest is on the scaling $W(q)\propto q^\alpha$ with the typical momentum exchange $q$ in a collision, as this sets the scaling of the diffusion constant with temperature according to $D(T)=\sum_q q^2  W(q)\propto T^{3 + \alpha }$. For spin-polarized electrons, scattering from low energy 
charge-excitations the Brownian particle follows the scaling $W(q)\propto q^2$ which implies for the diffusion constant in this case $D(T)\propto T^5$~\cite{TJK-PRB10,AZT-PRB11,MA-PRB12}. The latter leads to the mobility $\mu\propto1/T^4$ in agreement with earlier results~\cite{Neto-Fisher,Gangardt-Kamenev}. To gain intuition how this result may change upon adding spin-degree of freedom to the problem we first discuss the limit of weak interactions.

{\it Weakly interacting electrons}. We consider weakly interacting electrons described by the Hamiltonian 
\begin{equation}
H=\sum_{p\varsigma} \varepsilon_p c^\dagger_{p,\varsigma}c_{p,\varsigma} 
+ \frac{1}{2L}\sum_{q\varsigma} V_q \varrho_{q, \varsigma}\varrho_{-q, \varsigma} .
\end{equation}
Here we assume a quadratic spectrum $\varepsilon_p=p^2/2m$, where $m$ is the electron effective mass,  
 $\varrho_{q,\varsigma }=\sum_pc^\dagger_{p+q,\varsigma}c_{p,\varsigma}$
is the charge density of electrons with spin quantum number $\varsigma$, $V_q$ 
denotes the Fourier component of the interaction potential, and $L$ is the system size.
 
For weak interactions the physical mechanism of relaxation in the one-dimensional system
was attributed to three-particle collisions~\cite{RMM-PRL09,TJK-PRB10}. Kinematic considerations suggest that the required momentum transfer of $2p_F$ to backscatter from the left to the right Fermi point cannot be accommodated within a single three-particle collision. Rather, the momentum $2p_F$ is transferred within a sequence of three-particle scattering events accommodating small momentum transfer $\delta p\ll p_F$. In the course of such multi-stage scattering processes a hole passes through the bottom of the band between the two Fermi points at $\pm p_F$ and multiple particle-hole pairs are created at the Fermi level.
 
The kinetic equation for the distribution $F(p)$ of a hole in the vicinity of the bottom of the band again takes the form of a Fokker Planck equation \eqref{fp}. The diffusion constant \eqref{d} in this case is expressed in terms of the probability for scattering of a hole
\begin{equation}\label{Weak Int}
W(q)=\sum_{Q_2Q_3}\sum_{Q'_2Q'_3} \Lambda^{\varsigma\varsigma'}_{pp'}
f(p'_2)[1-f(p_{2})]f(p'_3)[1-f(p_{3})],
\end{equation}
where $\Lambda^{\varsigma\varsigma'}_{pp'}$ is the quantum mechanical rate for a three electron scattering process from initial states $I=\{ Q_1,Q_2,Q_3\}$ into final states $F=\{ Q'_1,Q'_2,Q'_3\}$ 
characterized by quantum numbers $Q_i=\{ p_i,\varsigma_i \}$ and correspondingly for $Q'_i$. The sum runs over all intermediate states involving the scattering of hole-state $Q_1=\{p,\varsigma\}$ 
into $Q_1'=\{p+q,\varsigma' \}$. Notice that upon linearizing the dispersion near the Fermi points $\varepsilon_p=\pm v_Fp$, Fermi distribution functions
$f(\pm p)=(e^{\pm v_Fp/T}+1)^{-1}$ and one can perform momentum summations exactly, $\sum_pf(p+k)[1-f(p)]=\frac{L}{2\pi}kn(k)$ and $\sum_p f(-p-k)[1-f(p)]=\frac{L}{2\pi}k[1+n(k)]$ with $n(k)=(e^{v_Fk/T}-1)^{-1}$ the Bose distribution function. Eq.~\eqref{Weak Int} then becomes structurally identical to Eq.~\eqref{d} as should be expected since the combinations of Fermi distributions $f(1-f)$ in~\eqref{Weak Int}  
describe particle-hole excitations which correspond to the bosonic modes that are emitted/absorbed in a scattering process~\eqref{d}. 

The central step in determining the diffusion constant is  the calculation of transition rates via Fermi's Golden rule $\Lambda^{\varsigma\varsigma'}_{pp'}=2\pi | \mathcal{A}^{\varsigma\varsigma'}_{pp'} |^2\delta( E_I-E_F)$, where $E_{I/F}$ labels energies of the initial and final states. Following then previous works~\cite{Lunde,AZT-PRB11} and carefully taking into account exchange contributions we find from second order perturbation theory \cite{supplementary}
\begin{equation}\label{a}
\mathcal{A}^{\varsigma\varsigma'}_{pp'}\propto 
\Xi^{\varsigma'_{1}\varsigma'_{2}\varsigma'_{3}}_{\varsigma_1\varsigma_2\varsigma_3}
\frac{V_{p_F}(V_{p_F}-V_{2p_F})}{\varepsilon_FL^2}\left(\frac{p_F}{q}\right)  \delta_{P_I P_F}.
\end{equation}
Here $\delta_{P_I P_F}$ ensures momentum conservation with $P_{I/F}$ being momenta of the initial and final states. The spin structure of the scattering rate is governed by the matrix 
$\Xi^{\varsigma'_{1}\varsigma'_{2}\varsigma'_{3}}_{\varsigma_1\varsigma_2\varsigma_3}=\delta_{\varsigma_1\varsigma'_{2}}
\delta_{\varsigma_2\varsigma'_{3}}\delta_{\varsigma_3\varsigma'_{1}}-\delta_{\varsigma_1\varsigma'_{3}}
\delta_{\varsigma_2\varsigma'_{1}}\delta_{\varsigma_3\varsigma'_{2}}$.
Crucially, we notice that this particular spin structure forces all spin-polarized contributions to compensate each other. Amplitudes involving different spin orientations and
spin-flips, however, remain singular in the transferred momentum. That is, for spin polarized electrons the leading contribution from the maximally exchanged terms cancels and the three-particle transition rate is dominated by subleading contributions in $q$, $\sum_{\varsigma\varsigma'}|\mathcal{A}^{\varsigma\varsigma'}_{pp'}|^2\propto \ln^2(p_F/|q|)$~\cite{AZT-PRB11}. On the other hand, taking into account spin degree of freedom gives $\sum_{\varsigma\varsigma'}|\mathcal{A}^{\varsigma\varsigma'}_{pp'}|^2\propto 1/q^2$. This suppression of the scattering amplitude in the spin-polarized case can be traced back to Pauli's exclusion principle in three-particle collisions~\cite{torsten,TA-1D-3p}. 

Building on the previous discussion, and noting that $p$-summations in Eq.~\eqref{Weak Int} give two extra powers in $q$, one finds  $D(T)\propto T^3$  in contrast to $D(T)\propto T^5$ in the spinless case. Correspondingly $\mu(T)\propto 1/T^2 $ and $\mu(T)\propto 1/T^4$ in the two cases, so that spin degree of freedom parametrically suppresses the mobility of a hole at the band bottom. A more detailed calculation  gives 
\begin{equation}\label{D-3pc}
D(T) \simeq \frac{V^2_{p_F}(V_{p_F}-V_{2p_F})^2}{v^4_F}
\left(\frac{T}{\varepsilon_F}\right)^3p^2_F\varepsilon_F
\end{equation} 
up to a numerical factor of order one~\cite{supplementary}.

{\it Spinon in a Luttinger liquid}. We proceed to study the fate of the above result beyond the weak interaction limit by including situations in which electrons fractionalize into spin and charge modes. To this end, we start out from a free model for the relevant excitations described by the Hamiltonian 
$H_0=H_0^d+H_0^\sigma + H_0^\rho$, 
where 
$H^d_0=\sum_\varsigma \varepsilon^\sigma_p d^\dagger_{p,\varsigma} d_{p,\varsigma}$
is the mobile free spinon in the vicinity of the band bottom Eq.~\eqref{disp}. $H_0^{\rho(\sigma)}$ are the standard Luttinger liquid Hamiltonians for bosonic charge- and spin-excitations described by the displacement fields $\phi_{\rho(\sigma)}$. The latter are conveniently expressed in terms of bosonic creation and annihilation operators 
\begin{equation}
\phi_l(x)=i\sum_q\sqrt{\frac{\pi K_l}{2 |q|}}e^{iqx-\eta |q|}
\left( b^{l \dagger}_{-q}+b^l_{q}\right),
\end{equation}
which diagonalize respective Hamiltonians 
$H^l_0=\sum_q \omega^l_q b^{l \dagger}_q b^l_q,\quad l=\rho,\sigma$.
In general, the dispersion of bosonic excitations is nonlinear and has an acoustic form $\omega^l_q=v_lq$ only at low momenta $q\to0$.
For repulsive interactions $v_\rho>v_\sigma$, and LL parameters obey $K_l=v_F/v_l$. 

To account for interactions between excitations we introduce couplings in the density-density and spin-spin channels, $H_1=H^\sigma_1+H^\rho_1$, whose structures are dictated by locality and spin-rotational symmetry
\begin{equation}\label{h-int}
H^\rho_1\!=\!\lambda_\rho\!\!  
\int\!\! dx\, \varrho(x) \varrho_d(x), \quad
H^\sigma_1\!=\! \lambda_\sigma\!\! \int\!\! dx\, \bold{S}(x) \bold{s}(x).
\end{equation}
Here  $\bold{s}= \sum_{\varsigma\varsigma'}\bm{\sigma}_{\varsigma\varsigma'}d^\dagger_\varsigma d_{\varsigma'}$ and $\varrho_d=\sum_\varsigma d_\varsigma^\dagger d_\varsigma$ are the spinon's spin- and particle-density, while $\lambda_{\rho(\sigma)}$ are respective coupling constants. The corresponding densities for bosonic excitations are conveniently 
expressed in terms of the fields $\phi_l$:
\begin{align}
\label{dens}
\varrho(x)&= -{\sqrt{2}\over \pi} \partial_x\phi_\rho(x), 
\quad
S^z(x)= -{\sqrt{2}\over \pi} \partial_x\phi_\sigma(x),
\\
S^x(x)&= {1\over \pi a} \cos[\phi_\sigma(x)],
\quad
S^y(x)= {1\over \pi a} \sin[\phi_\sigma(x)],
\end{align}
where $a\sim k^{-1}_{F}$ is the short distance cutoff. The above Hamiltonian is introduced based on the recently developed phenomenology~\cite{Schmidt-PRB10,Pereira-Sela-PRB10}. Coupling constants $\lambda_l$ can in principle be fixed microscopically employing Galilean invariance and SU(2) symmetry, but are treated as mere parameters of the model in the following. 

\begin{figure}[t]
	\centering
		\includegraphics[width=.49\textwidth]{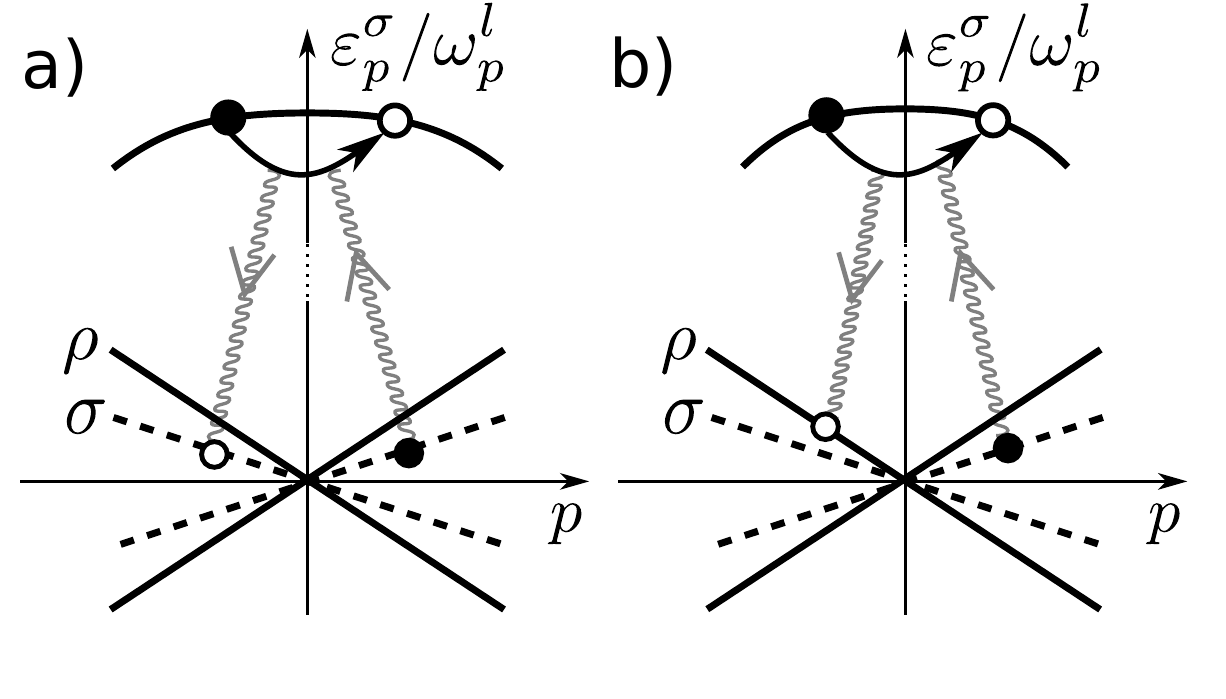}
	\caption{Schematics of two of the most relevant scattering processes contributing to the spinon's mobility. Panel a): A high energy spinon is backscattered absorbing and emitting a spin excitation of the LL. Panel b): A mixed backscattering process with the absorption of a spin and emission of a charge excitation.}
	\label{fig:spinon_scattering}
\end{figure}

{\it Transition rates}. We have prepared the stage for a calculation of the spinon diffusion coefficient \eqref{d} beyond the weak interaction limit. Generalizing the preceding analysis we study the rates $\Gamma^{\varsigma\varsigma'}_{ll'}$ of transition from an initial into a final spinon state accompanied 
by absorption and emission of bosonic spin or charge excitations, i.e. 
$| I \rangle= d^\dagger_{p,\varsigma} b^{l\dagger}_{k} |0\rangle$ and  
$\langle F| =\langle 0| d_{p+q,\varsigma'} b^{l'}_{k'}$, 
with $l,l'=\sigma,\rho$, see Fig.~\ref{fig:spinon_scattering}.

As in the weakly interacting limit, kinematic constraints enforce vanishing of scattering rates calculated in first order perturbation theory. Leading order contributions thus arise from the next, second order -- Raman scattering processes. To separate then contributions where spinons scatter off spin- and/or charge-excitations it is convenient to write $\Gamma^{\varsigma\varsigma'}_{ll'} =2\pi |M^{\varsigma\varsigma'}_{ll'}|^2 \delta(E_F-E_I)$, 
where 
\begin{equation}\label{2pt}
M^{\varsigma\varsigma'}_{ll'}= \left\langle F\left|{H^{ll'}_2 + H^{l'l}_2 \over 1 + \delta_{ll'}}\right|I\right\rangle,
\quad
H^{ll'}_2= H^{l'}_1 {1\over E_I - H_0} H^l_1
\end{equation}
and to discuss individual contributions separately. To this end, we bring the effective Hamiltonian into the form
\begin{align}\label{struct}
\hspace{-.126cm}
H_2^{l'l}
&=\sum_{\varsigma\varsigma',jj'} \sum_{pkk'}  
d^\dagger_{p+k+k',\varsigma}  
X^{l'j'}_{k'} 
{ h_{l'j';'lj}^{\varsigma\varsigma'} \over E_I - H_0-\varepsilon^\sigma_{p+k} }
X_{k}^{lj} 
d_{p,\varsigma'}
 \end{align}
where $X^{\sigma j}=S^j$ the $j=x,y,z$ components in the spin channel and $X^\rho=\varrho$ the charge  channel. We next discuss the main results for the different scattering channels.

{\it Scattering off charge excitations}. The scattering of a spinon by absorption and emission of a low-energy charge excitation is described by $h_{\rho\rho}^{\varsigma\varsigma'}\propto\lambda^2_\rho \delta_{\varsigma\varsigma'}$ in Eq.~\eqref{struct}. The corresponding transmission amplitude $M_{\rho \rho}^{\varsigma\varsigma'}$ can then be calculated straightforwardly. We leave aside calculational details, see Ref.~\cite{supplementary}, and note here only that upon inserting typical momenta dictated by kinematic constraints, one finds that $M^{\varsigma\varsigma'}_{\rho\rho} \propto (\lambda^2_\rho q/m^*v^2_\rho) \delta_{\varsigma\varsigma'}$. This, of course, just leads to the result $D(T)\propto T^5$ like for scattering of an impurity in a spinless Luttinger liquid~\cite{Neto-Fisher,Gangardt-Kamenev,Lamacraft}.

{\it Scattering off spin and charge excitations}. Scattering of a spinon accompanied by absorption and emission of spin-excitations is described by $h_{\sigma j, \sigma j'}^{\varsigma\varsigma'}\propto\lambda^2_\sigma \left( \delta_{jj'} \delta_{\varsigma\varsigma'}+  i\epsilon_{ijj'}\bm{\sigma}^i_{\varsigma\varsigma'} \right)$ in Eq.~\eqref{struct}, where $\epsilon_{ijk}$ is the Levi-Civita tensor. The first contribution in $h_{\sigma j, \sigma j'}^{\varsigma\varsigma'}$ preserves the projection value of spin and is structurally identically to $h_{\rho\rho}^{\varsigma\varsigma'}$. 
The second contribution in $h_{\sigma j, \sigma j'}^{\varsigma\varsigma'}$ is structurally different from the previous ones. As the products of operators $S^x$, $S^y$ and $S^x$, $S^z$ involve odd numbers of creation/annihilation operators these combinations do not contribute to the scattering rate of interest. This leaves us with scattering processes involving the product $S^y$, $S^z$, describing spin-flip events.
Further making use of normal ordering of the spin operators we may linearize $S^y=  \phi_\sigma/(\pi a)$. This allows to directly calculate corresponding transition amplitude~\cite{supplementary}, which turns out to be the leading one of all allowed scattering channels. Invoking kinematic constraints we find $M^{\varsigma\varsigma'}_{\sigma\sigma} \propto (\lambda^2_{\sigma} k_F/m^*v^2_{\sigma})\bm{\sigma}^y_{\varsigma\varsigma'}$, which implies that the weak interaction result $D(T)\propto T^3$ holds at arbitrary interaction strength. Technically, the difference between contributions $h^{\varsigma\varsigma'}_{\rho\rho}$ and $h^{\varsigma\varsigma'}_{\sigma\sigma}$  results from the non-commutativity of spin operators, which prevents a cancellation of the leading terms in the amplitude. 

Similarly, mixed processes involving spinon-scattering off charge- and spin-excitations, give to leading order $M^{\varsigma\varsigma'}_{\rho\sigma}\propto (\lambda_\rho\lambda_\sigma k_F/m^*v_\rho v_\sigma)\bm{\sigma}^y_{\varsigma\varsigma'}$. Notice that these again involve spin-flip processes, while those processes conserving the spin-projection value $M^{\varsigma\varsigma}_{\rho\sigma}$ are subleading in $q/k_F\ll1$. A careful calculation of all the relevant contributions \cite{supplementary} results in the diffusion coefficient  (again up to a numerical factor of order one)
\begin{equation}\label{result}
D(T) \simeq \left(\frac{\lambda_\sigma\sqrt{K_\sigma}}{v_\sigma}\right)^4
\left(\frac{T}{\varepsilon_\sigma}\right)^3
\left(1 + \frac{\lambda^2_\rho v^3_\sigma}{8\lambda^2_\sigma v^3_\rho}
\right)\varepsilon_\sigma p^2_F
\end{equation}
describing the scattering of a Brownian spinon from spin and charge excitations of a Luttinger liquid, where $\varepsilon_\sigma=m^*v^2_\sigma/2$. Eqs.~\eqref{D-3pc} and \eqref{result} are the main results of this paper. 

As mentioned above, in the case of antiferromagnetic coupling diffusion coefficient also scales with $T^3$, however the physics reason of this behavior is different and can be traced back to the two-channel Kondo problem \cite{Lamacraft}. For the strongly repulsive interactions of the Wigner crystal limit $D$ also scales with $T^3$ \cite{MAK-PRB14}. We conclude that independent of the interaction strength spin degrees of freedom suppress the mobility of large momentum excitations in one-dimensional quantum liquids in a parameter $\sim(T/\varepsilon_F)^2\ll1$. This includes situations where spin and charge decouple. 

{\it Discussion and Summary}. We have studied the diffusion coefficient of a spinon in a Luttinger liquid, and shown that the resulting mobility of the spinon is parametrically suppressed at low temperatures compared to that of a mobile excitation in the spinless case. The motion of an impurity in a quantum liquid is one of the central concepts of LL theory with applications e.g. to hole-dynamics in semiconducting nanowires or impurities in ultracold quantum gases. The diffusion coefficient discussed in this work sets the equilibration rate in generic Luttinger liquids as $\tau^{-1}\propto D(T) e^{-\Delta_\sigma/T}$. Relatedly, it also defines transport properties of clean quantum wires. Specifically, interaction induced backscattering process result in corrections to the quantized conductance $G=(e^2/\pi)(1 - \delta g)$ displaying activation behavior
 $\delta g \propto (LD/\sqrt{m^*T^3}) e^{-\Delta_\sigma/T}$~\cite{fn1}. An activated behavior of $\delta g$ has also been observed in the recent experiments Refs.~\cite{G-Exp-1,G-Exp-2,G-Exp-3}. Verification of the pre-exponential temperature-dependence $\delta g \propto T^{3/2} e^{-\Delta_\sigma/T}$ predicted here would provide an important test for our understanding of equilibration effects in clean quantum wires.

\textit{Acknowledgments}. We would like to thank K.A. Matveev, A.V. Andreev and A.D. Klironomos for numerous  discussions and for sharing results of their work~\cite{MAK-PRB14} prior to the publication. T.M. acknowledges support by Brazilian agencies CNPq and FAPERJ. Work by M.-T. R. was supported by the Alexander von Humboldt Foundation. This work at MSU (A.L.) was supported by NSF Grant DMR-1401908.

\end{document}